\documentstyle[aps,twocolumn,epsfig]{revtex}

\begin{document}

\title{Vortex formation in weakly-interacting inhomogeneous Bose condensates}

\author{B. Jackson, J. F. McCann, and C. S. Adams}
\address{Dept. of Physics, University of Durham, South Road, Durham, 
DH1 3LE, England.}

\date{\today}
\draft
\maketitle

\begin{abstract}

We solve the time-dependent Gross-Pitaevskii in 2-D to simulate the
flow of an object through a dilute Bose condensate
trapped in a harmonic well. We demonstrate vortex formation
and analyze the process in terms of the accumulation of phase-slip
and the evolution of the fluid velocity.

\end{abstract}



The origin of drag in quantum liquids is central to the
understanding of superfluidity \cite{nozi90}. For a weakly-interacting fluid,
Bogoliubov showed that macroscopic occupation of the ground state
leads to a linear dispersion curve, and hence superfluidity for
motion slower than a critical velocity. 
However, for liquid helium,
the transition to normal flow is observed at a much lower
velocity than can be explained by the dispersion curve. This led 
Feynman \cite{feyn55} to suggest that the onset of dissipation 
may arise due to vortex shedding, but experimental verification of 
this idea has been impeded by the two-fluid
nature of liquid helium, which complicates quantitative comparison with theory.\\

The recent experimental discovery of
Bose condensation in dilute alkali vapours \cite{ande95,davi95,brad97}
presents a near-perfect system for the study of quantum fluids: the condensates
are almost pure, and sufficiently
dilute that the interactions can be accurately parameterized in terms of
a scattering length. As a result, a relatively simple
non-linear Schr\"odinger equation (NLSE), known as the Gross-Pitaevskii equation
\cite{gpe}, gives a precise description of condensate dynamics.
Experiments have confirmed that the NLSE is remarkably 
accurate in the limit of low temperature \cite{jin96,mewe96,dodd97}.\\

Recent experiments have also demonstrated that far-off resonant laser light may be
used to split \cite{ande97} or excite the condensate \cite{ande97a}.
The question arises whether such light forces may be used to
study the phenomenon of vortex formation, and thereby shed
some light on the issue of superfluidity.
The appearance of vortices in solutions of the 2-D NLSE has
been considered by Frisch {\it et al.} \cite{fris92}. However, they
consider homogeneous fluid flow past an impenetrable obstacle, 
which is inappropriate for atomic condensates. 
Here, we solve the time-dependent 2-D NLSE to simulate the
flow of an `object' through a dilute inhomogeneous condensate. The `object' is the
potential barrier produced by far-off resonant blue-detuned laser beam 
\cite{davi95,ande97}.
The condensate is formed in a harmonic trap with the obstacle 
at the center. Subsequently, the obstacle is withdrawn at a constant speed, $v$
(from the Galilean invariance of the NLSE \cite{nozi90}, this
is equivalent to flow past a stationary object).
Vortices are formed when the condensate, whose motion is limited
by the speed of sound, cannot adjust sufficiently quickly to the
density wave induced by the object. We show how the time
scale for vortex formation can be extracted from the 
phase-slip of the wavefunction.\\

The evolution of the condensate is described by the NLSE \cite{units}:
\begin{equation}
i \partial_t \psi = \left(-\nabla^2 +V+C\vert\psi\vert ^2\right)\psi~,
\end{equation} 
where $C= 8 \pi N a $ is the nonlinear coefficient, 
$N$ is the number of atoms in the condensate,
and $a$ is the $s$-wave scattering length. The potential
term, $ V= {1 \over 4} (x^2 +y^2) + \alpha
\exp (- \beta x^2 -\beta (y -v t)^2 ) $, describes a symmetric harmonic trap
with a light-induced `Gaussian' potential barrier, moving with velocity, $v$. 
We have considered a variety of obstacle parameters but present detailed results
for $C=500$, $\alpha=30$, and $\beta=3$.\\

\begin{figure}
\epsfig{file=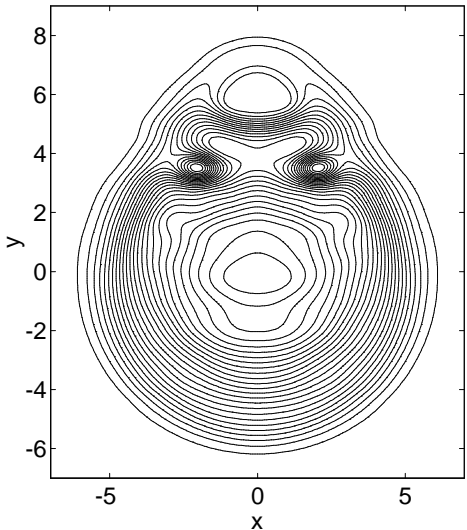,clip=,width=7.5cm,bbllx=120,bblly=170,bburx=540,bbury=575 }
\caption{Contour plot of the condensate density, $|\psi(x,y,3.0)|^2$ 
for $v=2.0$. There are 25 contours, equally spaced between 0.0 and 0.0182.}
\label{fig:1}
\end{figure}

\begin{figure}
\epsfig{file=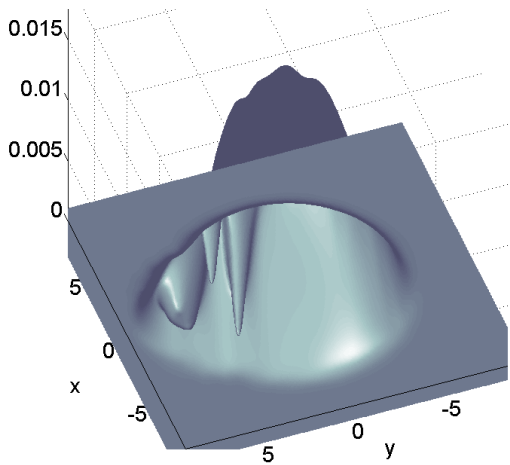,clip=,width=7.5cm,bbllx=140,bblly=200,bburx=500,bbury=500 }
\caption{Surface plot of the condensate density, $|\psi(x,y,3.0)|^2$ 
for $v=2.0$, viewed from below, illustrating the minima corresponding to the object
and the vortex-pair.}
\label{fig:2}
\end{figure}

The eigenvalue problem, $\psi=\phi (x,y) e^{-i \mu t}$, 
(where $\mu$ is the chemical potential), and $v=0$,
is easily solved using finite difference methods.
With $\alpha=0$, we find $\mu=9.003$. The depletion of density in the
trap center due to the presence of the obstacle, (with $\alpha=30$, $\beta=3$,
$v=0$), raises the chemical potential to $\mu=9.208$. A change
in the object potential creates a local disturbance  
which propagates through the fluid. The time evolution 
of $\psi$ was determined by the split-operator technique \cite{numerics}. 
Extinction of the object,
excites sound waves \cite{ande97a}, 
which propagate with speed, $c=\sqrt{2C|\psi|^2}$, ($\sim 4.3$ at the peak density).
Translation of the object displaces the condensate center-of-mass \cite{com}, 
and leads to vortex-pair creation as illustrated in Fig.~\ref{fig:1}. 
The vortex minima are highlighted by the surface plot shown in 
Fig.~\ref{fig:2}. The half-width of 
the vortex core is comparable to the healing length,
$\lambda \sim C^{-1/4}$. We have simulated the free
expansion of the condensate and observe that
the vortices also expand, permitting 
experimental detection using optical imaging.\\

For low object speeds, no vortices are observed implying
that the process requires motion faster than a critical speed. For an 
impenetrable cylinder of diameter $d$, 
the critical speed is related to the energy required to create a vortex-pair,
$\epsilon=(C/2a) \ln (d/\lambda)$ \cite{nozi90}. 
For a finite potential barrier, we find that the vortices emerge from a
point, i.e., initially $d\sim \lambda$, so the pair creation energy, $\epsilon \sim 0$,   
and the critical speed depends principally on the
speed of sound. However, due to the condensate 
inhomogeneity, the speed of sound and hence the critical speed are spatially
dependent. For this reason, we choose to study the process of vortex formation
in terms of the fluid velocity and the phase of the wavefunction.\\

\begin{figure}
\epsfig{file=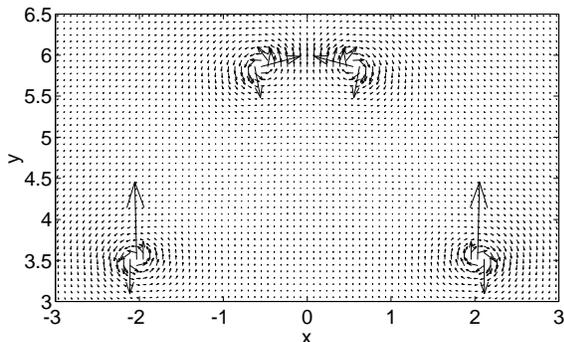,clip=,width=7.5cm,
bbllx=70,bblly=230,bburx=570,bbury=530 }
\caption{Plot showing the fluid velocity, $\mbox{\boldmath $v_s$}$, 
in the vicinity of the object, for $t=3.0$ and $v=2.0$. 
The arrow length is 40$v_{\rm s}$ in h.o.u.}
\label{fig:3}
\end{figure}

Fig.~\ref{fig:3} shows a quiver plot of the fluid velocity,
$\mbox{\boldmath $v_s$} = (\psi^* \mbox{\boldmath $\nabla$} \psi-
\psi \mbox{\boldmath $\nabla$} \psi^*)/i\vert\psi\vert^2$,
in the vicinity of
the object, for $t=3.0$ and $v=2.0$ (as in Figs.~\ref{fig:1} and \ref{fig:2}). 
One sees that vortices are formed in pairs
with opposing vorticity, and that the 
velocity is inversely proportional to distance from the
vortex line, as expected. Two pairs are discernable in Fig.~\ref{fig:3}: 
one at $y=3.5$, which corresponds to
the density zeros in Fig.~\ref{fig:2}, and a second at $y=5.9$, (not visible
in Fig.~\ref{fig:2}, because it has yet to separate from
the object). The second pair appears due to the rapid accumulation of phase-slip
at the edge of the condensate, as discussed below.\\

By calculating the wavefunction phase, $S(x,y,t)$, one may
verify that the phase changes by $2\pi$
on circulation of the vortex line \cite{nozi90}.
In addition, the evolution of the phase determines the time-scale for vortex formation. 
At $t=0.0$, the phase is uniform, then the motion induces a dephasing
or `phase-slip' centred on the object. 
The phase-slip between two neighbouring points along the
$y$-axis is defined as: $\Delta S(y,t) \equiv {\rm arg} 
\left\{\psi(0,y+\Delta y,t)\right\}-{\rm arg}\left\{ \psi(0,y,t)\right\}$, 
with $ -\pi < {\rm arg}\left\{z\right\} \leq \pi$.
The maximum phase-slip, $\Delta S_{\rm max}$, 
occurs at a $y$-coordinate close to the center of the
object.\\ 

\begin{figure}
\epsfig{file=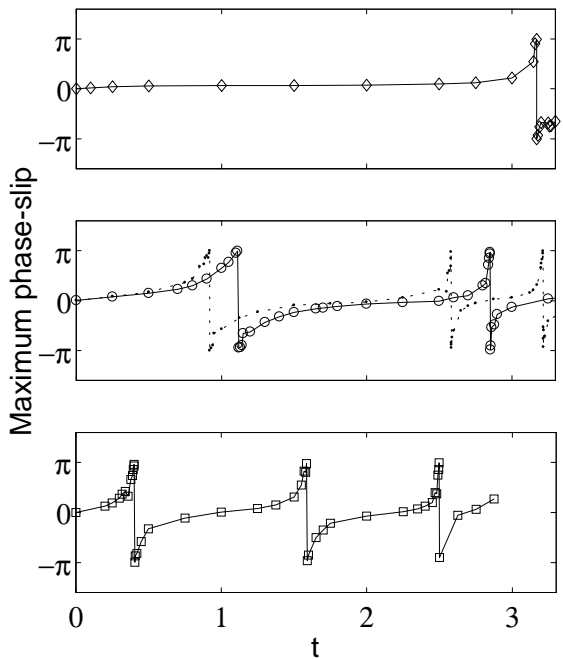,clip=,width=7.5cm,bbllx=145,bblly=170,bburx=485,bbury=580 }
\caption{Time-evolution of the maximum phase-slip, $\Delta S_{\rm max}$, 
for $v=1.3$ (top); $v=2.0$
(middle); and $v=3.0$ (bottom). For $v=2.0$ we show an additional
curve for $C=200$ (dotted), illustrating that vortices are produced faster
when the speed of sound is slower.}
\label{fig:4}
\end{figure}

Fig.~\ref{fig:4} shows a plot 
$\Delta S_{\rm max}$ for different object speeds.
For $v=2.0$ and $3.0$, the phase-slip accumulates gradually, 
reaching a value of $\pi$, and then changing sign at a critical time, $t_{\rm c}$.
The sign change indicates a reversal of the flow, and 
may be taken to define the moment of vortex creation.
The critical time is proportional to the speed of sound:
vortices are produced faster for smaller $c$, as the condensate
requires longer to adjust to the perturbation (compare curves for $C=200$ and
$C=500$ in Fig.~\ref{fig:4}). 
For $t>t_{\rm c}$, the phase-slip builds again until another vortex-pair is
formed. For $v=1.3$, (upper plot in Fig.~\ref{fig:4})
the phase-slip initially saturates and then increases towards the edge
of the condensate, where the density
and hence the local speed of sound are lower. This explains the sudden
formation of a vortex-pair as the object leaves the condensate, (e.g.
the second pair for $v=2.0$ appearing in Figs.~\ref{fig:3}
and \ref{fig:4}). For $v=3.0$, a third pair is created at $t=2.5$:
the object is still surrounded by condensate at this time because of
a motion induced `stretching'.\\

\begin{figure}
\epsfig{file=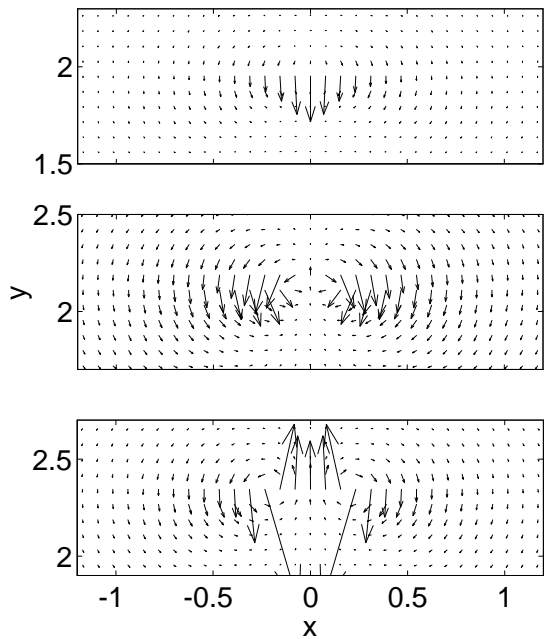,clip=,width=7.5cm,
bbllx=140,bblly=175,bburx=490,bbury=575 }
\caption{Plot showing the velocity field in the vicinity of the object
for $v=2.0$, at times spanning the instant of 
vortex formation: $t=1.0$ (top), 1.1 (middle), and 1.2 (bottom). 
The arrow length is 10$v_{\rm s}$ in h.o.u.}
\label{fig:5}
\end{figure}

The evolution of the velocity field while the vortex is forming
is shown in Fig.~\ref{fig:5}. For $t=1.0$ ($t<t_{\rm c})$, the on-axis flow is
backwards, i.e., opposite to the direction of motion, as the fluid attempts
to fill the `hole' left by the departing object. At $t=1.1$, the fluid 
begins to skirt
around the object and fill the hole with a forward flow. 
The wavefunction nodes (separated by less than one grid point at this stage)
are pulled apart due to the gradient in the object potential. 
By $t=1.2$, their separation is comparable to the healing length, and 
the pattern of vortex flow is beginning to emerge. For $t>1.2$ the vortex pair
continue to separate, and the on-axis forward flow gradually decreases,
reverting to a reverse flow at $t=2.0$ (i.e., when the phase slip changes
sign, see Fig.~\ref{fig:4}). Thereafter, the process repeats. 
For longer times, the motion of the vortex
pair is sensitive to
the exact parameters used, and a detailed study is required
before any generalizations can be made.\\

To summarize, we have solved the time-dependent NLSE in 2-D to simulate the
flow of an object through a dilute Bose condensate
trapped in a harmonic well. The process of vortex formation
has been analyzed in terms of the accumulation of phase-slip and
the evolution of the velocity field.
We find that the vortices emerge from a point
close to the center of the object, and that the 
vortex shedding frequency is higher when the speed of sound is lower.\\

A complete description of dilute atomic vapours should include
the effects of the non-condensate density, 
finite temperature, and dissipation. To what extent the 
NLSE provides an accurate description of
vortex formation requires quantitative
comparison between simulation and experiment. Such a comparison 
will provide the focus for future work.

\acknowledgements
Financial support was provided by the Nuffield Foundation and  the EPSRC.\\


\begin{references}

\bibitem{nozi90}
P. Nozi\`eres and D. Pines, {\it The Theory of Quantum Liquids}, Vol. II,
(Addison-Wesley, Redwood City, 1990).

\bibitem{feyn55}
R. P. Feynman, Prog. Low Temp. Phys. {\bf 1}, 17 (1955).

\bibitem{ande95}
M. H. Anderson, J. R. Ensher, M. R. Matthews, C. E. Wieman,
and E. A. Cornell, Science {\bf 269}, 198 (1995).

\bibitem{davi95}
K. B. Davis, M. O. Mewes, M. R. Andrews, N. J. van Druten, D. S. Durfee,
D. M. Kurn, and W. Ketterle, Phys. Rev. Lett. {\bf 75}, 3969 (1995).

\bibitem{brad97}
C. C. Bradley, C. A. Sackett, and R. G. Hulet, Phys. Rev. Lett.
{\bf 78}, 985 (1997).


\bibitem{gpe} 
V. L. Ginzburg and L. P. Pitaevskii, Sov. Phys. JETP {\bf 7},
858 (1958); E. P. Gross, J. Math Phys. {\bf 4}, 195 (1963). 

\bibitem{jin96}
D. S. Jin, J. R. Ensher, M. R. Matthews, C. E. Wieman,
and E. A. Cornell, Phys. Rev. Lett. {\bf 77}, 420 (1996).

\bibitem{mewe96}
M.-O. Mewes, M. R. Andrews, N. J. van Druten, D. M. Kurn, D. S. Durfee, 
C. G. Townsend, and W. Ketterle, Phys. Rev. Lett. {\bf 77}, 988 (1996).

\bibitem{dodd97}
R. J. Dodd, M. Edwards, C. W. Clark, and K. Burnett, preprint.

\bibitem{ande97}
M. R. Andrews, C. G. Townsend, H.-J. Miesner, D. S. Durfee, 
D. M. Kurn, and W. Ketterle, Science {\bf 275}, 637 (1997).

\bibitem{ande97a}
M. R. Andrews, D. M. Kurn, H.-J. Miesner, D. S. Durfee, C. G. Townsend,
S. Inouye, and W. Ketterle, Phys. Rev. Lett. {\bf 79}, 547 (1997).

\bibitem{fris92}
T. Frisch, Y. Pomeau, and S. Rica, Phys. Rev. Lett. {\bf 69}, 1644 (1992).

\bibitem{units}
We use scaled harmonic oscillator units (h.o.u.) throughout, i.e., 
for a symmetric harmonic trap with angular frequency
$\omega$ and particles of mass $m$, the units of length, time and
energy are $( \hbar / 2 m \omega)^{1/2}$, $\omega^{-1}$ and $\hbar \omega$,
respectively. 

\bibitem{numerics}
Most of the calculations were performed using a
square box of side 10 h.o.u. divided into a grid
of 256$\times$256 points. A larger grid (20 h.o.u., 512$\times$512) was used
to check for edge-effects. More 
details of the numerical methods employed are discussed 
elsewhere: B. Jackson, J. F. McCann, and C. S. Adams, to be published.

\bibitem{com}
The center-of-mass motion provides a useful diagnostic:
the initial acceleration is $v$ divided by the integration time-step;
after a few steps the solution is only sensitive to $v$; and 
when the object leaves the condensate, the center-of-mass
oscillates at the trap frequency.

\end{references}
\end{document}